# An Emergency System for Succoring Children using Mobile GIS

Ayad Ghany Ismaeel
Department of Information Systems Engineering-
Erbil Technical College-Technical Education Foundation
Erbil-Iraq

***Abstract*— The large numbers of sick children in different diseases are very dreaded, and when there isn't succor at the proper time and in the type the sick child need it that makes us lose child. This paper suggested an emergency system for succoring sick child locally when he required that, and there isn't someone knows his disease. The proposed system is the first tracking system works online (24 hour in the day) but only when the sick children requiring the help using mobile GIS. In, this emergency system the child will send SMS (for easy he click one button) contains his ID and coordinates (Longitude and Latitude) via GPRS network to the web server (the child was registered previously on that server), in this step the server will locate the sick child on Google map and retrieve the child's information from the database which saved this information in registration stage, and base on these information will send succoring facility and at the same time informing the hospital, his parents, doctor, etc. about that emergency case of the child using the SMS mode through GPRS network again. The design and implement of the proposed system shows more effective cost than other systems because it used a minimum configuration (hardware and software) and works in economic mode.**

## I. Introduction

Increasing the rating of sick children in different diseases per year base on World Health Organization WHO and UNICEF, e.g. over the last 40 years asthma particularly in children approximately 300 million people worldwide currently have asthma and its prevalence increases by 50% every decade, in North America, 10% of the population has asthma [1].

Other disease is congenital heart defect over 1,000,000 babies born with this disease worldwide each year 100000 of them will die within the first year, i.e. one of every 100 infants they have congenital heart defect to some extent [2]. There are 70,000 children (less than 1 year -14 years) worldwide are expected to develop type 1 diabetes annually per year increase is estimated at around 3% [3].

From what advancement the sociality face to face a front of a big problem with this large numbers of sick children from view of three important diseases only, i.e. what about other diseases this problem become more effective if there isn't who offer succoring and helping at a suitable time for those children when they are in school, shopping, with their friends, etc, i.e. will lose a large number from those children if they needed succoring and there isn't someone with the child known which type and suitable help he needed.

To solve this problem sure will think about new track system for those sick children to succor them, not like these traditional tracking systems which are developed so far use a handheld GPS receiver device for tracking the location depend on real time tracking and continuity on the interval of tracking [4] for example, of these types of traditional tracking systems; may construct from n-tier as shown in Fig. 1 [6].

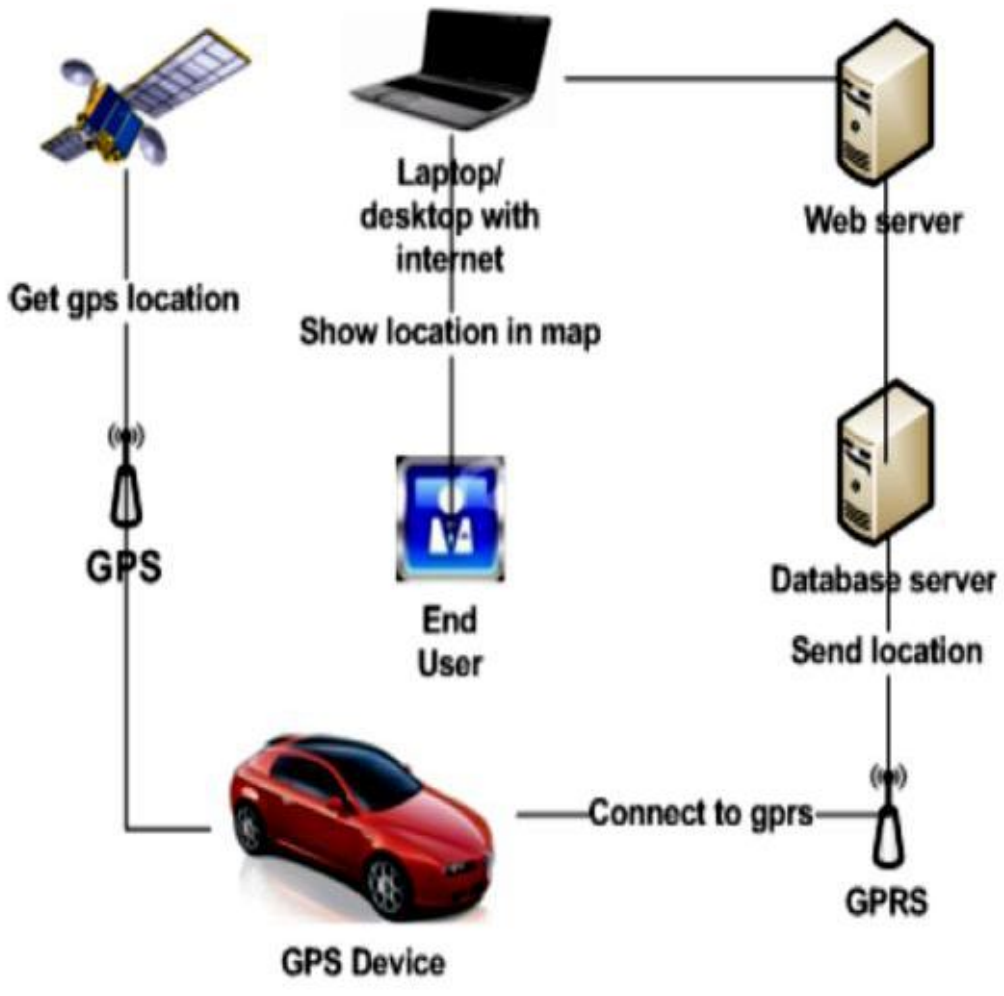

Figure 1. The N-Tier Tracking System Diagram [6].

Really there is needed for an emergency system can tracking the sick children only when they required help and there are no one knows what is the disease of each of them, at a suitable time and in the quality which the children are required using new techniques and modes to satisfy system in effective cost.

## II. Related Work

Katina Michael and others [2006] employed usability context analyses to draw out the emerging ethical concerns facing current human-centric GPS applications personal locators for children, the elderly or those suffering from loss, and monitoring of parolees for law enforcement, security or personal protection purposes. The outcome of the study is the classification of the current state GPS applications into the contexts of control, convenience, and care; and a preliminary ethical framework for considering the viability of GPS location-based services emphasizing privacy, accuracy, property and accessibility [4].

Alahakone, A.U. and Veera Ragavan [2009] presented the development of a geospatial information system for path planning and navigation of mobile objects, The system involves a GIS implemented using Google maps to visualize the routes of mobile objects acquired from GPS receivers over a GPRS network [5].

Ruchika and BVR [2011] proposed a cost effective method of tracking a human's mobility using two technologies via GPRS and GPS, and further the cost is reduced by using GPRS rather than using SMS for communicating the information to the server, but the tracking system is design based on Android only, not for any mobile phone (general) can support GPS and General Packet Radio Service GPRS like iPhone, windows phone, iPad, etc [6].

The whole systems allow the user's mobility to be tracked using a mobile phone which is equipped with an internal GPS receiver and a GPRS transmitter, i.e. most of the applications developed so far use a handheld GPS receiver device for tracking the location [6], and real time tracking and continuity on the interval of tracking may be very high cost specially when the server, IP network, and ISP are busy in the interval of tracking [4].

To overcome the problems above, an emergency system must be contain the following techniques and modes:

*A. Mobile GIS:*

As expansion of GIS technology from the office into the field, a mobile GIS enables field-base personnel to capture, store, update, manipulate, analyze, and display geographic information. Mobile GIS integrates one or more of the following technologies:

1) Mobile devices.
2) Global Position System (GPS).
3) Wireless communications for Internet GIS access.

There is wide using of mobile GIS to complete the multiple tasks one of them the tracking for persons, vehicle, etc [7].

*B. An Emergency System Works Via GPRS:*

GPRS is the widely acknowledged successful application to adopt; it cannot be denied that high costs are involved in both setting up as well as maintenance of the application. Also, in order to full integrate and fully tap upon the efficiency of the system, and

*C. Short Message Service SMS:*

The mobile terminal sends data through SMS to the receiving terminal, compare to the modem solution the SMS solution is more economical because the tracking system will work in an emergency cases only (when really the child needs the succoring and help), i.e. to overcome the time of tracking systems in general, which are used to maintain long time continual tracking system it would therefore, result again in a high cost in maintaining a continual tracking system.

## III. The Motivating

The objective of this paper reach to an emergency tracking system offers succoring (when the children needed the help), i.e. make the system works really only when the sick child required the succoring and in economic mode that will reduce the time and delay as well as the cost, to satisfy this aim must be thinking about new techniques and modes, so the proposed of an emergency system will involve the following characteristics:

*A. The Mobile GIS Technique:*

The system involves a GIS implemented using Google maps to visualize the location of mobile or track device for a sick child without needing to use GPS receiver, but the proposed system will base on supporting of a mobile build-in GPS technique on devices like iPhone, Windows phone, iPad, etc.

*B. Modes Of Transmission As Follow:*

1) *GPRS mode:* The mobile terminal sends data through GPRS data channel to a special TCP/IP server linked to the Internet, or a PC with a fixed Internet IP address. In this case, the GPRS is always online and billed only on the bytes transmitted, rendering it to be a much cheaper alternative to any current systems, and

2) *SMS mode:* The mobile terminal sends data through SMS to the receiving terminal, compared to the modem solution; the SMS solution is more economical.

*C. Server of TCP/IP mode:*

Have a fixed IP address, as well as a dynamic IP address as soon as the receiving server or receiving terminal get its IP address on boot up, the user will need only to reconfigure the IP address setting of the mobile terminal to align it to any of the users desired output this remote setting mechanism makes all this possible without any hassle.

## IV. Architecture of Proposed an Emeregncy System

The main tasks for the suggested design of an emergency system to succor the sick children summarize in the flowchart as show in Fig. 2.

The architecture of suggested an emergency system involves multiple modules as shown in Fig. 3, these modules are:

*A. Registration module:*

This module referring to any sick child needed to be serve using this an emergency system must be register via web interface constructed for the system one time only (no duplicate) and saved the child's information in a database created for registration. Without registration, this system can't recognize the child which needs succor/help.

*GIS module:*

In this module the sick child will use his mobile (a child who is equal or less than 15 year), this mobile can support GPS technique (built-in) to locate his coordinates (Latitude and Longitude) or using tracking device like GlobalSat TR-203 (a child who can't use mobile). Every time the child need succoring from this an emergency system will send SMS contains only the coordinates of location and child ID (number of child's mobile or his sequence number in database, etc) which specified in registration module (A above) using GPRS network, and this SMS will be received by the web server of an emergency system.

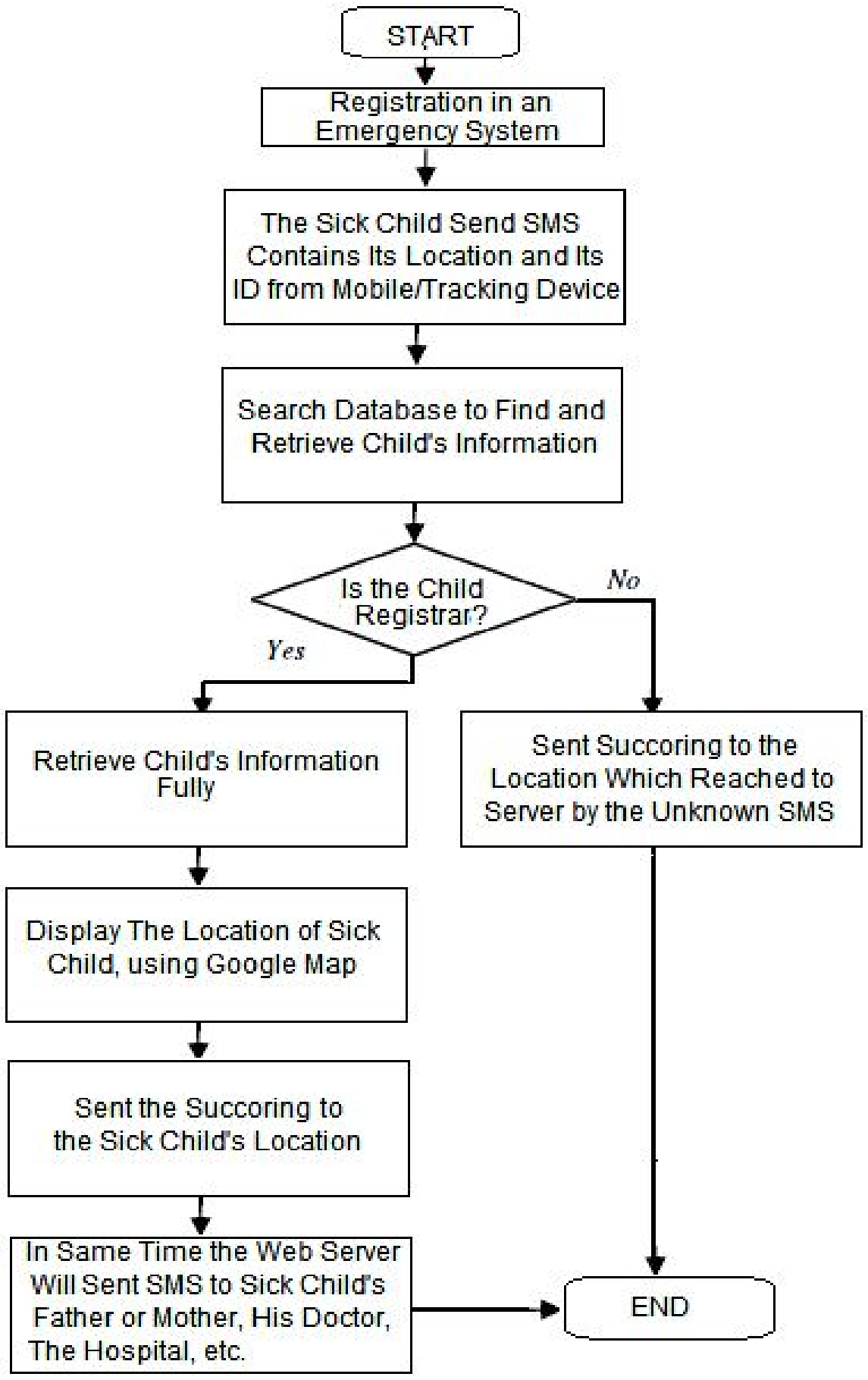

Figure 2. Flowchart for the main tasks of proposed an emergency system.

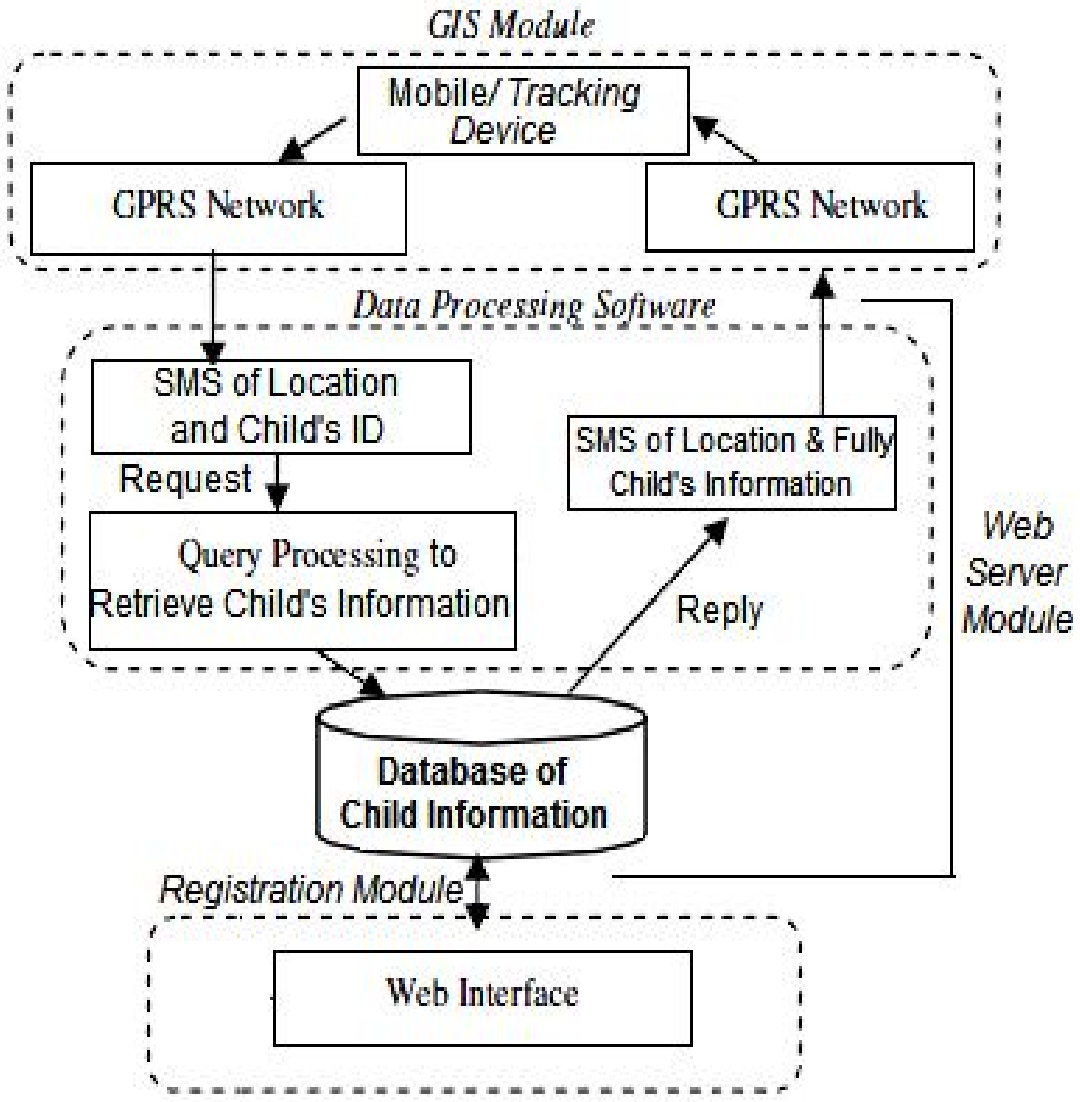

Figure 3. Architecture of proposed an emergency system.

## C. Web server module:

That module will work when an emergency system received SMS, the web server will use the Child ID within received SMS to search the database to find and retrieve the fully information of sick child, then send succoring facility (Car, Helicopter, Lifeboat, etc), and at the same time send SMS to informing the emergency hospital, the father or mother of sick child, etc.

Fig. 4 shows a diagram of serving or supporting sick child which is done in two stages by the proposed an emergency system, the first stage start when the child request the succoring/help from an emergency system, i.e. the serving start base on request of sick child, so not like a natural tracking system which was worked in continuity, while the proposed an emergency system offers serve or support for a sick child (at any time and online), but the real serve start when SMS of request reached from child to the web server as shown in Fig. 4; a.

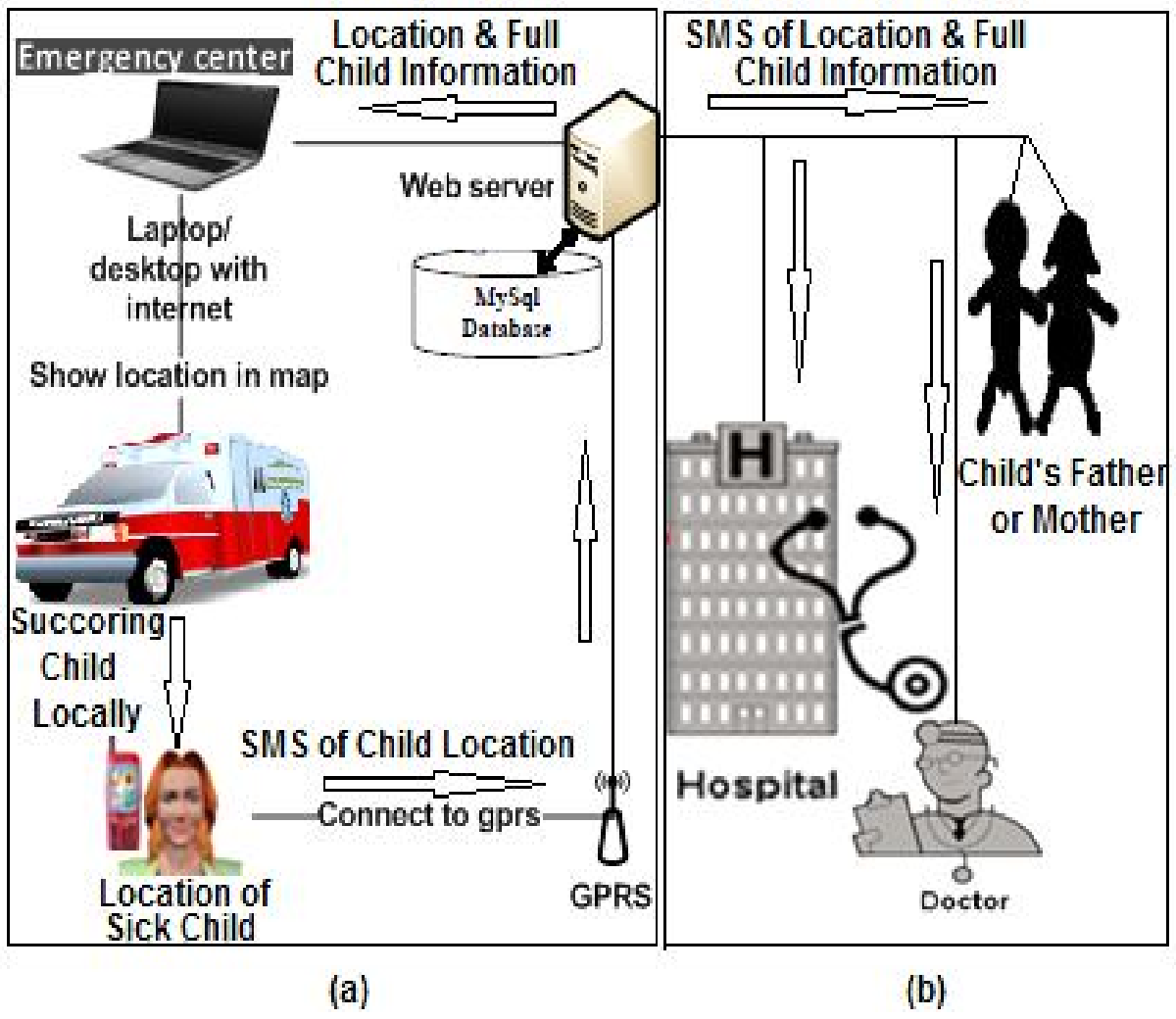

Figure 4. Serving/Supporting diagram of proposed an emergency system.

This feature will reduce and minimize the cost because the proposed system will need only web server based on a database not two servers one for web server and another for database, again this feature (non continuity of emergency system) will use SMS solution which is more economical because an emergency system works when the child needed not like the traditional tracking systems which were continuous, Fig. 4; b shows the second stage of serving an emergency system starting when retrieve the fully information of childlike name, type of diseases, father or mother, his doctor, etc the technique of finding and retrieving the child's information written in the flowchart as shown in Fig. 5.

# V. Experimental Results

Implementing the proposed design of an emergency system reveals below:

## A. The requirement of configuration

The configuration for suggested system in this research can be divided into:

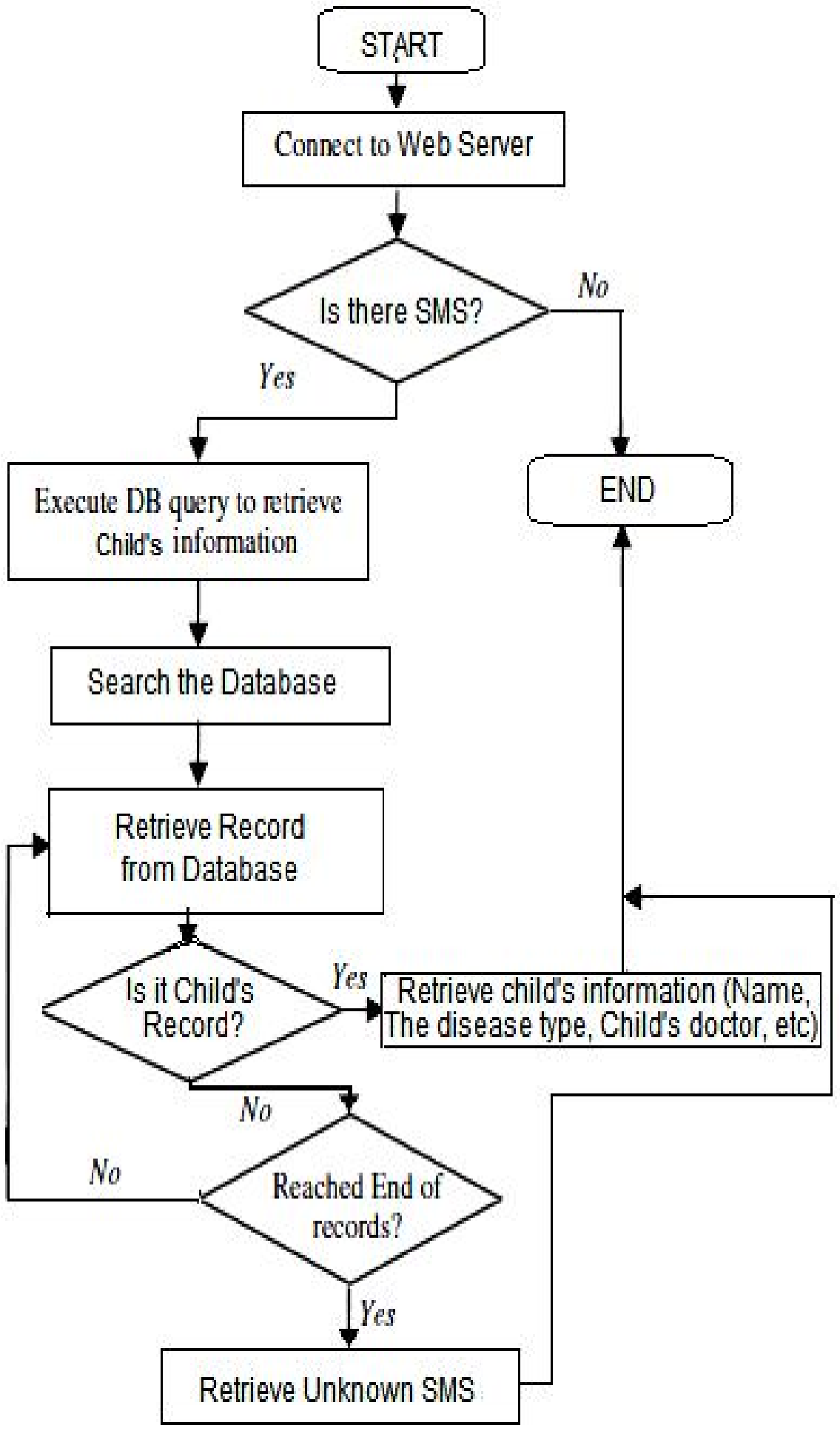

Figure 5. Flowchart technique of finding and retrieving child's information.

1) *The Software:* Tools which are needed windows server 2008 to setup and install web server, the other important software C# as programming language under Visual studio 2010 package which is used in the web interface for registration and for implement the all other techniques like the connection between the web server and database, then the search (find and retrieve) technique information of sick child, etc. the last software needed MySQL package use to construct the database involve table call info contains the fields as shown in Table 1.

TABLE I. FIELDS OF INFO TABLE

| Field | Type |
|---|---|
| ChildId | Int(10) |
| Name | Varchar(20) |
| Age | Varchar(7) |
| Father-No | Int(10) |
| Mother-No | Int(10) |
| Disease-Name | Varchar(20) |

2) *The Hardware:* As the first device will need a server friendly with windows and Microsoft packages selected type HP server, second need a mobile can support GPS technique (built-in), here selected windows phone which is friendly with Microsoft packages (the compatibility make avoid conflict) and cheap comparing to iPhone, iPad and tracking device (which can support GPS/GSM/GPRS technology like GlobalSat TR-203) as shown in Table 2. Finally, will need a resource to connect with Internet Service Provider ISP cross GPRS network.

TABLE III. COMPARING THE WINDOWS PHONE WITH OTHER TYPES OF MOBILES AND TRACKING DEVICE

| Feature | Windows Phone | iPhone | Tracking Device (TR-203) |
|---|---|---|---|
| Cost | Cheap | Expensive | Relatively Expensive |
| Zone | Unlimited | Unlimited | Limited |
| Multiple usage | Yes | Yes | No |

### *B. Implement the proposed System*

For An emergency system, which called Succor will select environment for implementation Erbil city and the first step are the registration of the child using GUI (web interface) from any PC connected to internet or from his mobile directly as shown in Fig. 6.

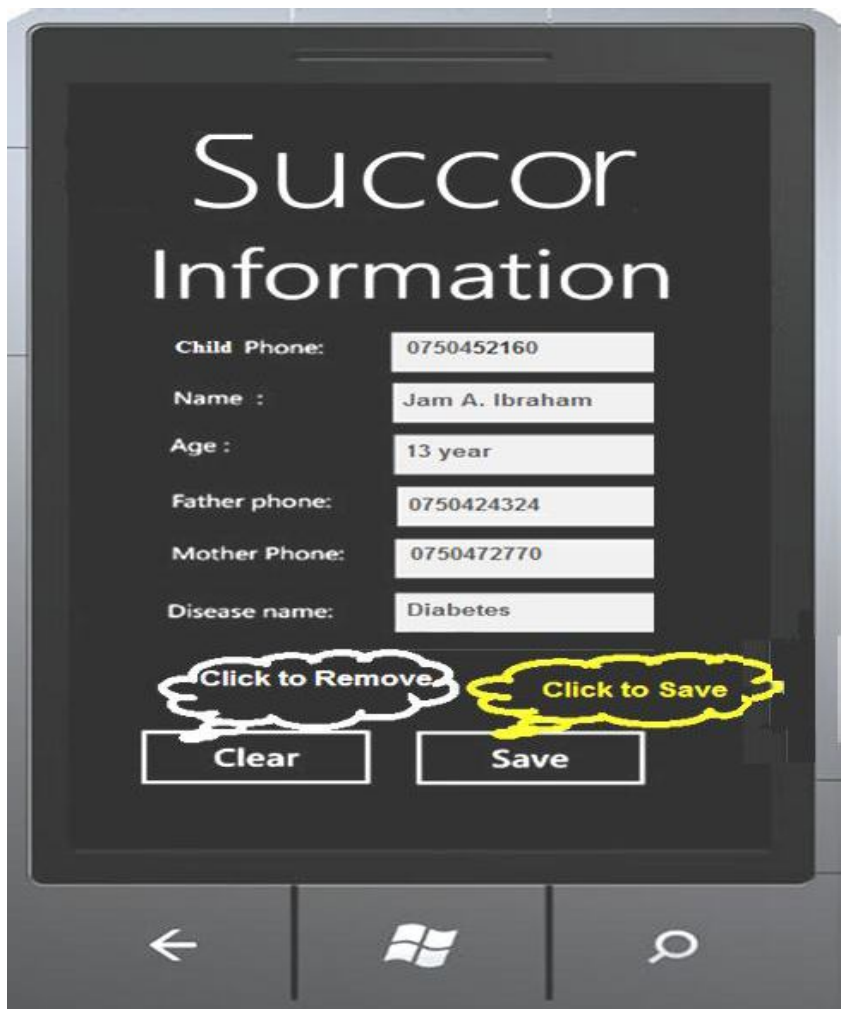

Figure 6. Registration of sick child in Emergency system using his mobile.

After registration the Succor system can offers serving/support to the sick child when request that by click one button (e.g. help button) and the location is determined automatically (Longitude and Latitude) by windows phone which support GPS functions as shown in Fig. 7 then the mobile will send SMS containing the coordinates and ChildId (phone number) to the web server.

The web server must be done a sequence of tasks as shown in Fig. 8:

1) *Search:* The web server will search the SMSs of requiring a succouring online/automatically or manually by click search button and these SMSs appear in the description location (see Fig.8).

2) *Find:* this button for finding fully information using ChildId (as key for finding) from the database of Succor system and appearing at information location as shown in Fig.

8, at the same time will determine the real location of sick child on Google maps.

*3) Send SMS:* the web server after sending succoring facility (e.g. car, helicopter, etc) to the real location of child, this button will use to send SMSs for father/mother of child (or any other persons from child's Family), and also to the emergency hospital (to become ready for receiving child), see Fig. 8.

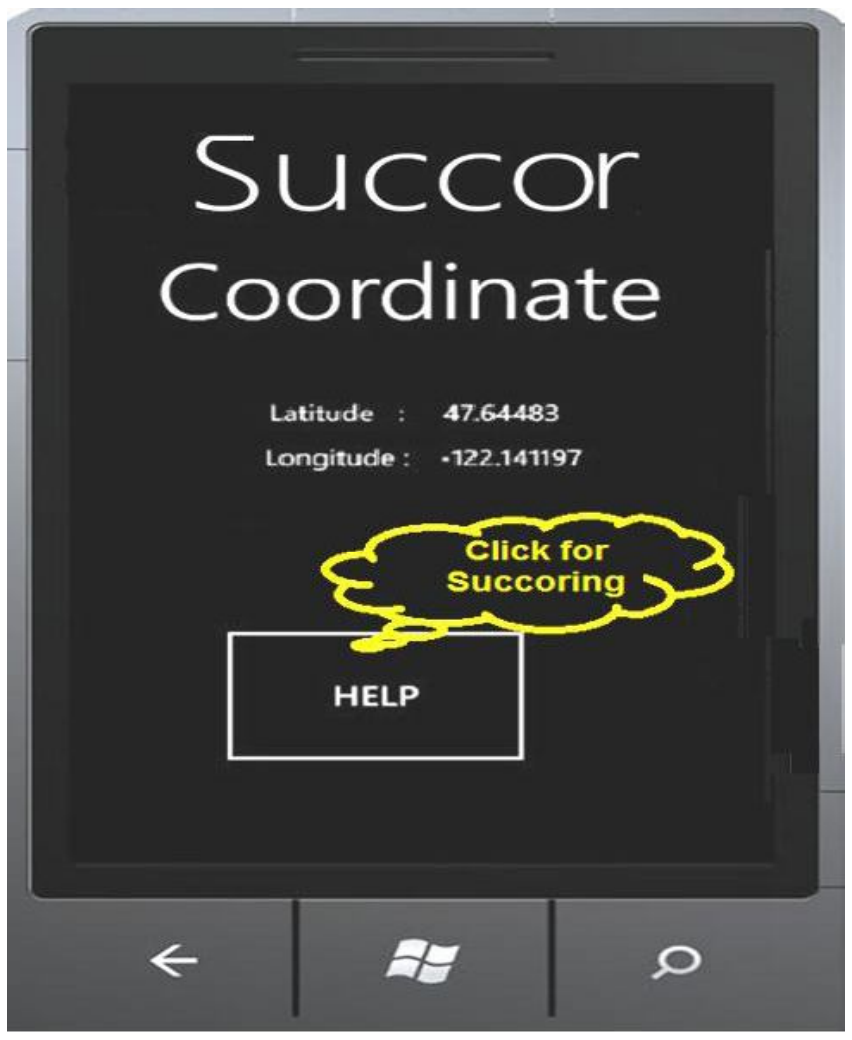

Figure 7. The sick child when need serve from succour (emergency) system.

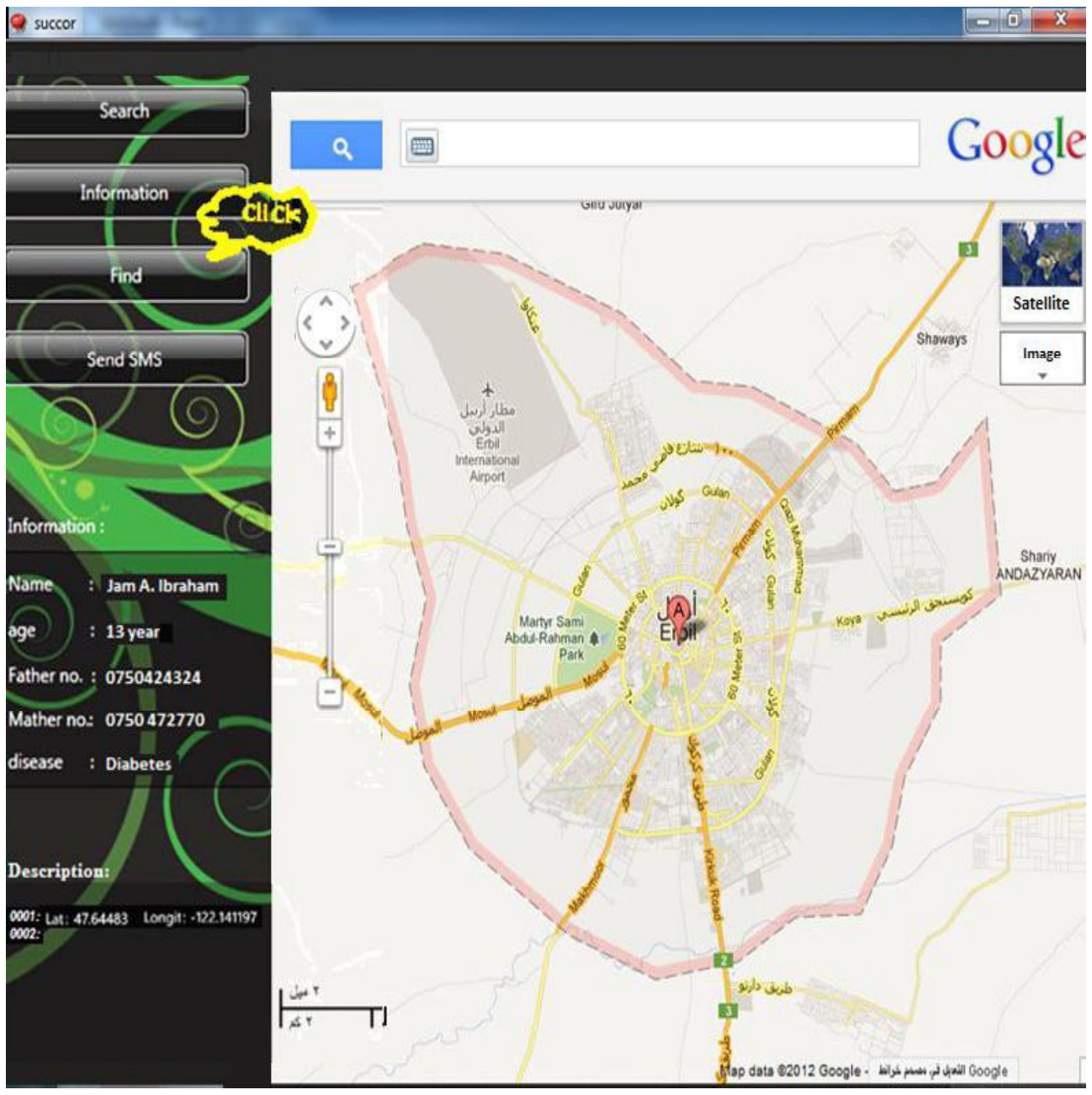

Figure 8. The tasks/buttons of web server of succouring system.

## C. *Discussion of the results:*

The important results of the proposed an emergency system comparing with other systems shown in Table 3.

TABLE IIIII. COMPARING THE PROPOSED SYSTEM WITH OTHER SYSTEMS

| **Feature** | **Proposed an Emergency System** | **Ruchika and BVR System** | **Alahakone and Veera System** |
|---|---|---|---|
| Request of Hardware | Relatively minimum request, e.g. no need extra GSM network, separated database server, etc | Relative middle request | Relatively maximum request |
| Serving continuity or at the request | At the request however, the serving is online 24 hours | Continuity | Continuity |
| Using SMS | Yes, it becomes economic with noncontinuity | No | Yes |
| Need the support of GPS receiver | No, because it supported by build-in GPS | Yes | Yes |
| Using specific mobile | No (General), which are support GPS | Yes | No |

# V. CONCLUSIONS AND FUTURE WORK

## A. *Conclusions*

The conclusions which are obtained from proposed system summarized as follow:

*1)* The design of proposed an emergency system is an effective cost more than other systems as referring to in Table 2; especially the proposed system is better than Ruchika and BVR system which is characterized based on cost effective [6].

*2)* The proposed system really works when the sick child send SMS, i.e. an emergency system is not continuity however, it works online and can accept any request at any time. This approach of succoring which is used the SMS mode relatively economic from other modes like modem solution.

*3)* This an emergency system can use as tracker for a child or any other type of ages whose make the registration or not in case of unregistered will send the succoring facility without any others information to the location from it the server received the SMS.

## B. *Future work*

Can extend the proposed system to achieve the follow tasks:

*1)* Make the an emergency system can do more GIS computing to determine the real location and distance (base on an update image satellite for the zone or location), this distance computed between an emergence center (which send the succoring facility) and the location of sick child as well as between the child's location and emergency hospital which take care of child, and then base on real location and the optimal path (distance) will decide to send a suitable type of succoring facility (helicopter, car, lifeboat, etc) and in the shortest path because selecting the suitable succoring facility and in shortest path are playing an important role in succoring the sick child.

Involve an emergency system knowledge about all emergency hospitals (e.g. by extended or added database) in the zone or location to inform the succoring facility to take the correct direction to the suitable hospital (proper to the type of the child's disease) to serve and help the child in a suitable an emergency hospital and at the shortest time (without any loss of time in this emergency case of child).

## ACKNOWLEDGMENT

Thanks to the team members (my students), who worked on my idea (project of succor group), and I was their mentor in competition of Imagine Cup for Microsoft in 2012.

## AUTHOR PROFILE

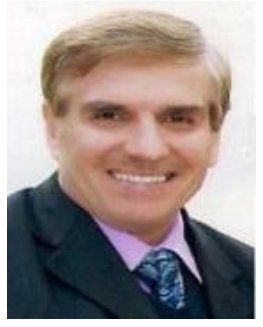

**Ayad Ghany Ismaeel** received MSC in computer science from the National Center of Computers NCC- Institute of Postgraduate Studies, Baghdad-Iraq at 1987, and Ph.D. computer science in qualification of computer and IP network from University of Technology, Baghdad- Iraq at 2006.

He is professor assistant, at 2003 and currently in department of Information Systems Engineering in Erbil Technical College-Iraq, His research interest in mobile, IP networks, Web application, GPS, GIS techniques, distributed systems and distributed databases. He is lecturer in postgraduate of few universities in MSC and Ph.D. courses in computer science and software engineering from 2007 till now in Kurdistan-Region, IRAQ.

Ayad Ghany Ismaeel is Editorial Board Member at International Journal of Distributed and Parallel Systems IJDPS http://airccse.org/journal/ijdps/editorial.html, Program Committee Member of conferences related to AIRCC worldwide, and reviewer in IJCNC (which is listed as per the Australian ARC journal ranking http://www.arc.gov.au/era/era_2012/era_journal_list.htm), IJDPS, IJCSIT journals (http://airccse.org/journal.html), and Conference CCSEIT-2012 within http://airccse.org/, as well as he adviser and reviewer in multiple national journals. The last published papers were in International Journal Distributed and Parallel System (IJDPS) as follow:

*GPS AND GIS, published in (IJDPS) Vol.3, No.2, March 2012. http://airccse.org/journal/ijdps/papers/0312ijdps05.pdf; NEW METHOD OF MEASURING TCP PERFORMANCE OF IP NETWORK USING BIO-COMPUTING, Published In (IJDPS) Vol.3, No.3, May 2012. Http://Airccse.Org/Journal/Ijd*